\documentclass{article}
\usepackage{breckenridge}
\usepackage{graphicx}
\frompage{000} \topage{000}                                              
\def\rightmark{Resonance Production in STAR}
\def\leftmark{C. Markert for the STAR collaboration}

\title{Resonance Production in STAR}
\authors{
{Christina Markert$^1$ for the STAR Collaboration}\\[2.812mm]
{\normalsize
\hspace*{-8pt}$^1$ Yale University, New Haven, CT 06520, USA\\[0.2ex]
\hspace*{-8pt} 
}}

 \abstract{The recent results from resonance production in
central Au+Au and p+p collisions at $\sqrt{s_{\rm NN}} = $ 200 GeV
from the STAR experiment at RHIC are presented and discussed.}

\keyword{resonances, lifetime, rescattering} \PACS{25.75.Dw}

\begin{document}

\maketitle \setcounter{page}{1}
\section{Introduction}
\label{intro}

The short lifetimes of resonances, a few fm/c, are of the same
order of the lifetime of the fireball source in a heavy ion
collision at RHIC. Measurements of the yields, widths, mass and
momentum distributions of resonances can provide information about
hadronization and the time span between chemical and thermal
freeze out. The resonances that we are currently investigating in
the STAR experiment are $\rho$ (1.3 fm/c), $\Delta$ (1.7 fm/c),
K*(892) (4 fm/c), $\Sigma$*(1385) (6 fm/c), $\Lambda$(1520) (13
fm/c) and $\phi$ (40 fm/c) (listed in table~\ref{resotable}).
Comparison of these resonances in heavy ion collisions with
elementary p+p and e$^{+}$+e$^{-}$ collisions may provide evidence
for possible medium effects in the extended volume of a heavy ion
reaction. By varying the centrality (impact parameter) and beam
species in heavy ion collisions, we can investigate the effect of
the fireball medium on the resonances.

\section{Analysis}
\label{ana}

The resonances are reconstructed from their charged decay
daughters (see table~\ref{resotable}) identified via energy loss
({\em dE/dx}) and their measured momenta in the {\em Time Projection Chamber} (TPC).
The resonance signal is obtained by the invariant mass reconstruction
of each daughter combination and subtraction of the combinatorial
background calculated by mixed event or like-signed techniques.
The resonance ratios, spectra and yields are measured at
mid-rapidity. The central trigger selection for Au+Au collisions
takes 7\% of the most central inelastic interactions. The setup
for the proton+proton interaction is a minimum bias trigger.

\begin{table}[htb]
\vspace*{-12pt} \caption[]{Resonances from PDG \cite{pdg98}}
\label{tab1} \vspace*{-14pt}
\begin{center}
\begin{tabular}{lllll}
\hline\\[-10pt]
Particle & mass & width & lifetime & decay channel \\
 &(MeV) & (MeV) &(fm/c) & \\
\hline\\[-10pt]
$\rho$(770) & 771.1 $\pm$ 0.9 &  149.2 $\pm$ 0.7 & 1.32 & $\pi$ +
$\pi$  \\
$\Delta$(1232) & 1232 $\pm$ 2 &  120 $\pm$ 5 & 1.64 & p + $\pi$ \\
\hline
K(892)$^{0}$ & 896.1 $\pm$ 0.27 &  50.7 $\pm$ 0.6 & 3.89 & K + $\pi$ \\
$\phi$(1020) & 1019.5 $\pm$ 0.02 &  4.26 $\pm$ 0.05 & 46.2 & K + K \\
$\Sigma$(1385)$^{+}$ & 1382.8 $\pm$ 0.4 &  35.8 $\pm$ 0.8 & 5.5 & $\Lambda$($\rightarrow$p+$\pi$) + $\pi^{+}$ \\
$\Sigma$(1385)$^{-}$ & 1387.2 $\pm$ 0.5 &  39.4 $\pm$ 2.1 & 5.0 & $\Lambda$($\rightarrow$p+$\pi$) + $\pi^{-}$\\
$\Lambda$(1520) & 1519.5 $\pm$ 1.0 &  15.6 $\pm$ 1.0 & 12.6 & p + K \\

\hline
\end{tabular}
\end{center}
\label{resotable}
\end{table}

\section{Resonances in Medium}\label{resopp}

During the expansion of the hot and dense fireball of a heavy ion
collision resonances are produced and a fraction of them decay due
to their short lifetimes inside the fireball medium. Two freeze
out conditions during such a fireball expansion can be
characterized by the end of the inelastic interactions (chemical
freeze out) and the end of elastic interactions (kinetic freeze
out). Elastic interactions of particles in the surrounding medium
(mostly pions) with resonances and their decay daughters can have
different effects. Elastic interactions with the resonances can
change their phase space distribution, for example the transverse
momentum distribution. Meanwhile elastic interactions with the
decay particles can result in a signal loss from invariant mass
reconstruction. Regeneration of the resonances can increase the
yield and also change the momentum distribution, because particles
forming a new resonance came from the medium. Resonances that
decay mostly into particles that are not pions will have a smaller
contribution from regeneration. The time between chemical and
thermal freeze out can be verified by comparing resonances with
different lifetimes. Transport model calculations maybe able to
describe medium effects at a microscopic level
\cite{ble02,ble02b}. These calculations were developed recently,
since heavy ion experiments are capable to measure resonances with
short lifetimes. The initial motivation for the measurement of
resonances was mass shifts and width broadenings of the mass
signal due to an influence of the medium (recent papers
\cite{lut01,lut02}). This signal can be "washed out" during the
expansion of the fireball source, when inelastic and elastic
interactions have an additional effect on the resonances and their
decay particles. Furthermore a width broadening would result in a
shorter lifetime. This would mean a higher probability of signal
loss for the hadronic decay channels due to rescattering of the
decay particles. The measurement from leptonic decay channels
would not have this effect. Therefore it is very important for our
understanding to measure the resonances through both hadronic and
leptonic decay channels.

\section{New Measurements}\label{new}

\begin{figure}[htb]
\centering
\includegraphics[width=0.47\textwidth]{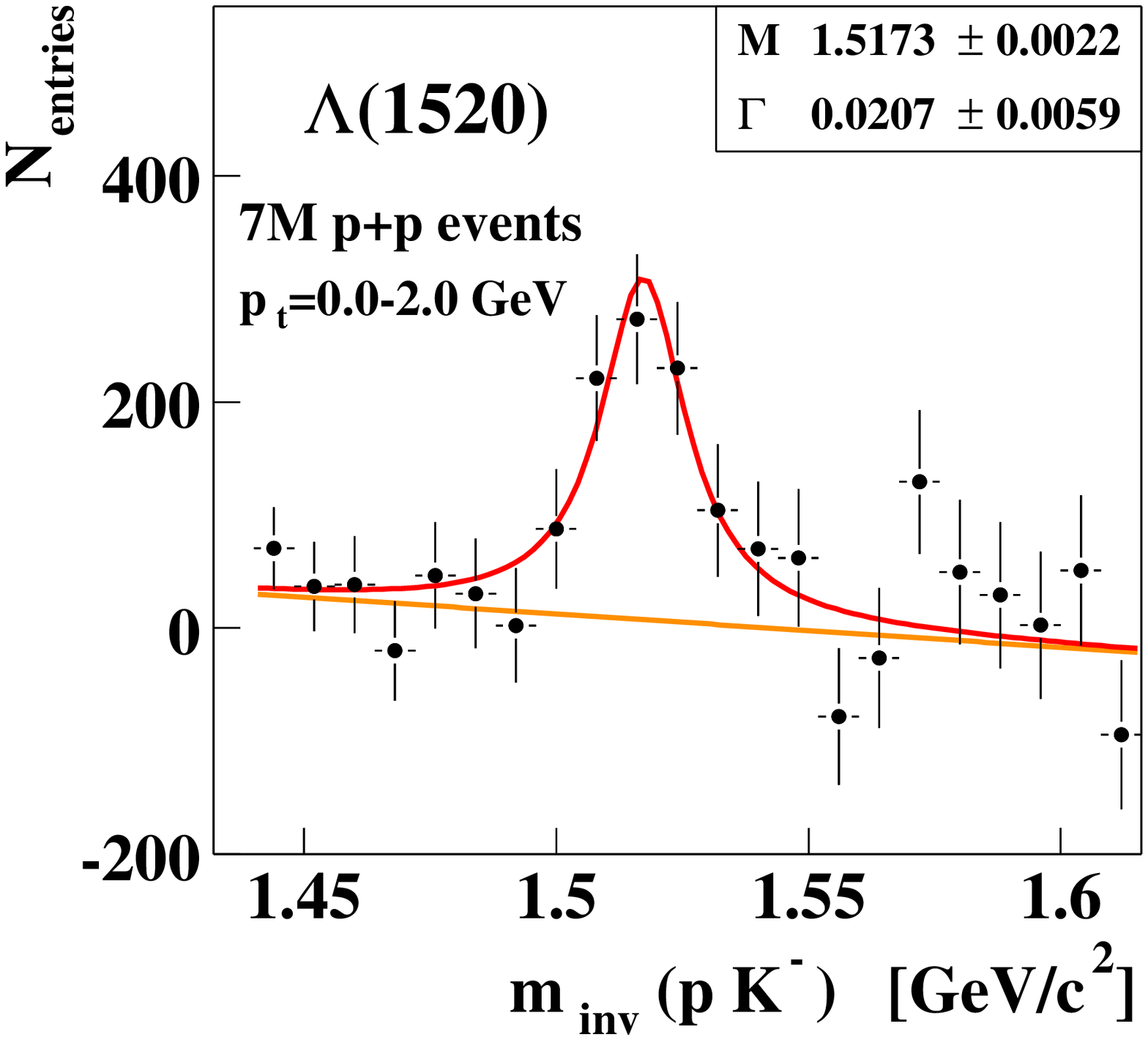}
\includegraphics[width=0.47\textwidth]{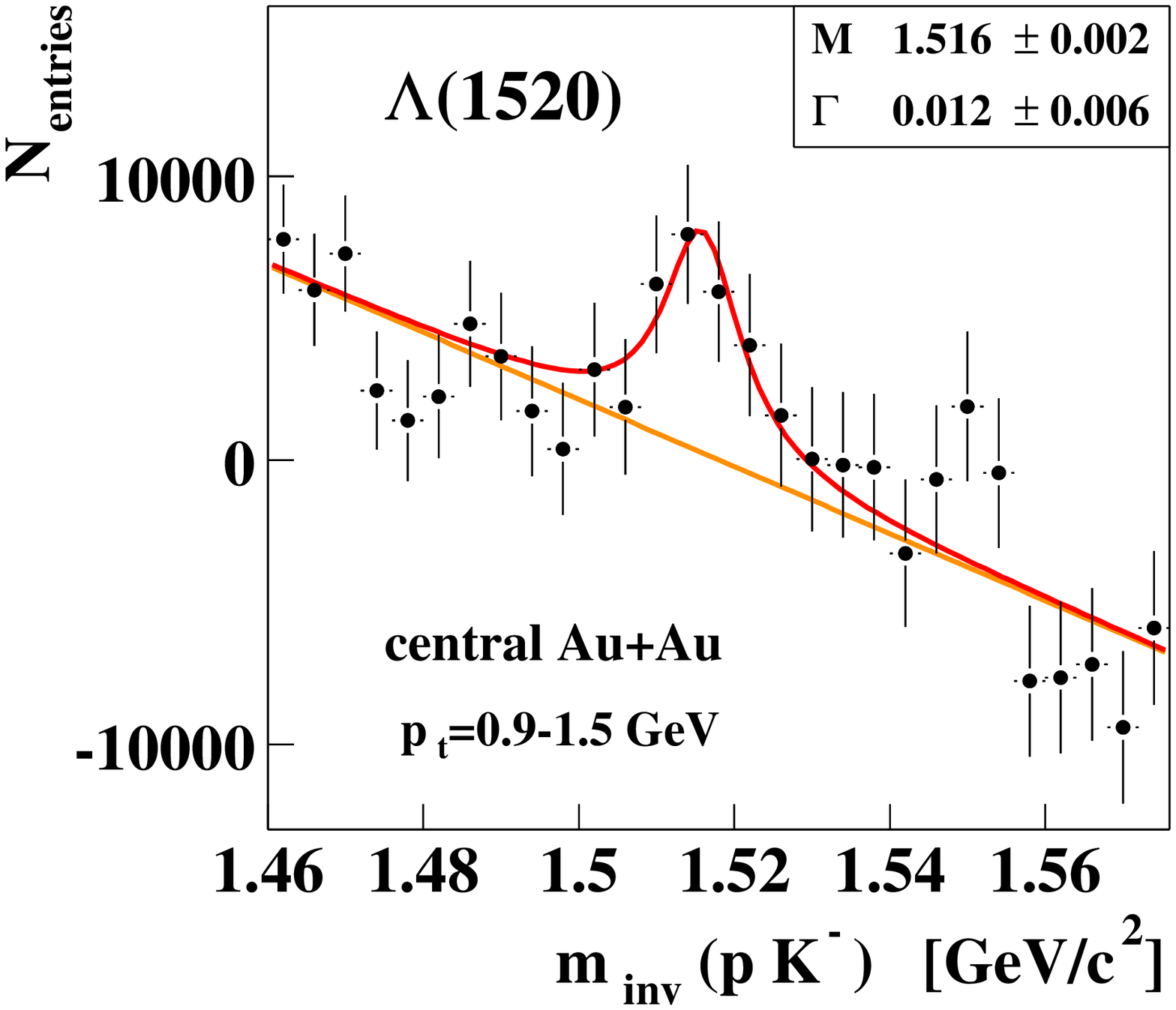}
\vspace{-0.7cm}
 \caption{Invariant mass distribution of p+$K^{-}$ in p+p (left) and in central
 Au+Au (right) collisions after mixed-event background subtraction. Left: Breit-Wigner-Fit with mass
1517~$\pm$~2~MeV/c$^{2}$ and width = 20~$\pm$~5 MeV/c$^{2}$. Right:
Breit-Wigner-Fit with mass = 1516~$\pm$~2~MeV/c$^{2}$ and width =
12~$\pm$~6 MeV/c$^{2}$ (from simulation 18~$\pm$~2 MeV/c$^{2}$ for nominal width + detector resolution) .}
 \label{lambda1520}
\vspace{-0.5cm}
\end{figure}

With the data set from the $\sqrt{s_{\rm NN}}=$200 GeV run of p+p
and Au+Au collisions STAR has sufficient statistics to measure the
$\Lambda$(1520) resonance for the first time.
Figure~\ref{lambda1520} shows the invariant mass signal of the
$\Lambda$(1520) after mixed event background subtraction, fitted
with a Breit-Wigner function plus a line fit in p+p (left) and
central Au+Au (right) collisions. The p$_{t}$ range is chosen in
respect to the best significance for the $\Lambda$(1520) signal
(p+p: p$_{t}$=0.0-2.0 GeV/c Au+Au: p$_{t}$=0.9-1.5 GeV/c). Due to
the low background for the p+p data we receive the best signal by
integration over the full the p$_{t}$ range. The extracted
mid-rapidity yield for p+p is
$\langle$$\Lambda$(1520)$\rangle$$_{|y|<0.5}$~=~0.0041~$\pm$~0.0006(stat)~$\pm$~10\%(sys).
For cental Au+Au, with an assumption of a inverse slope parameter
T=350~MeV, the measured yields is
$\langle$$\Lambda$(1520)$\rangle$$_{|y|<0.5}$~=~0.058~$\pm$~0.021(stat)~$\pm$~30\%(sys).
 \vspace{-0.25cm}

\vspace{-0.25cm}

\section{Meson Resonances}\label{data}

\begin{figure}[htb]
\centering
\includegraphics[width=0.48\textwidth]{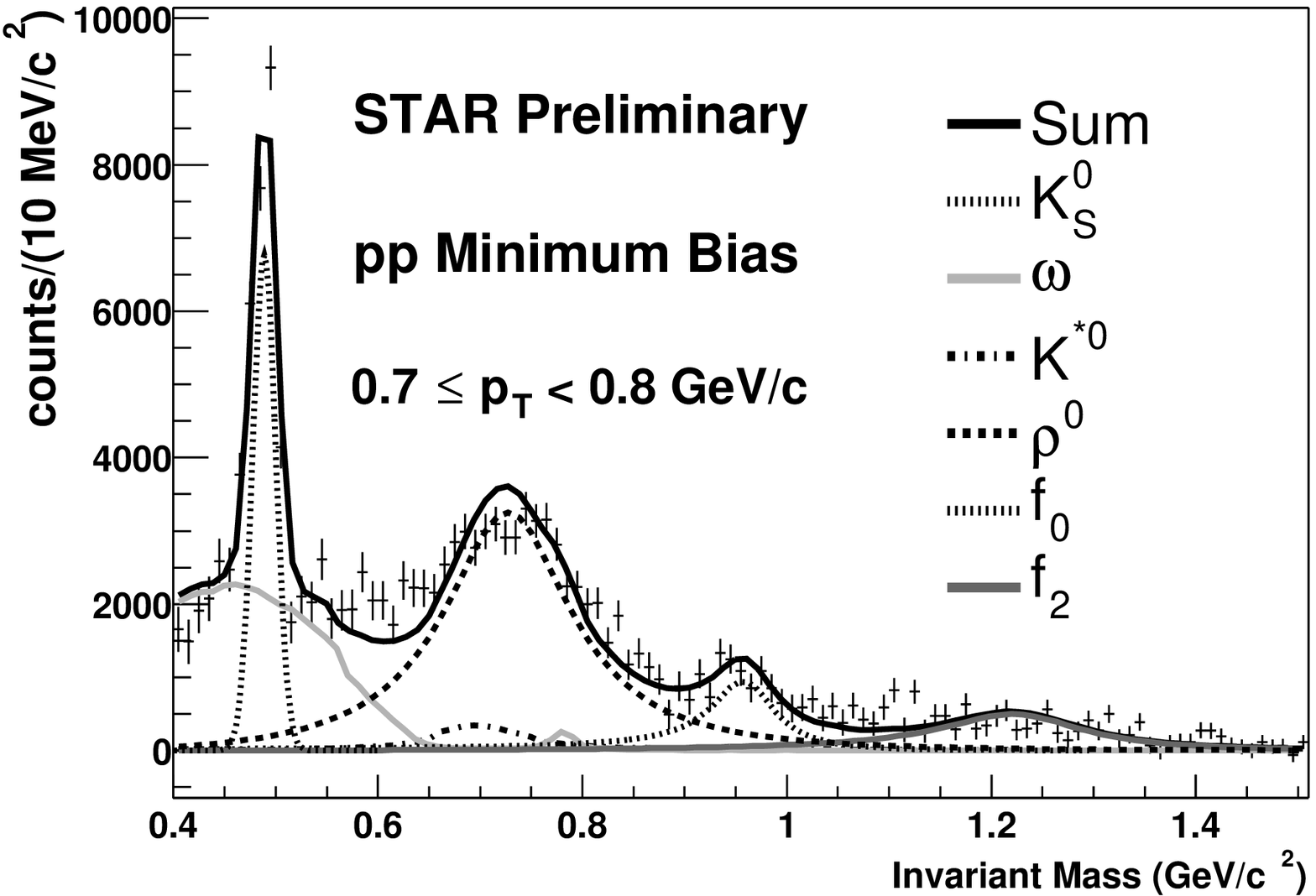}
\includegraphics[width=0.48\textwidth]{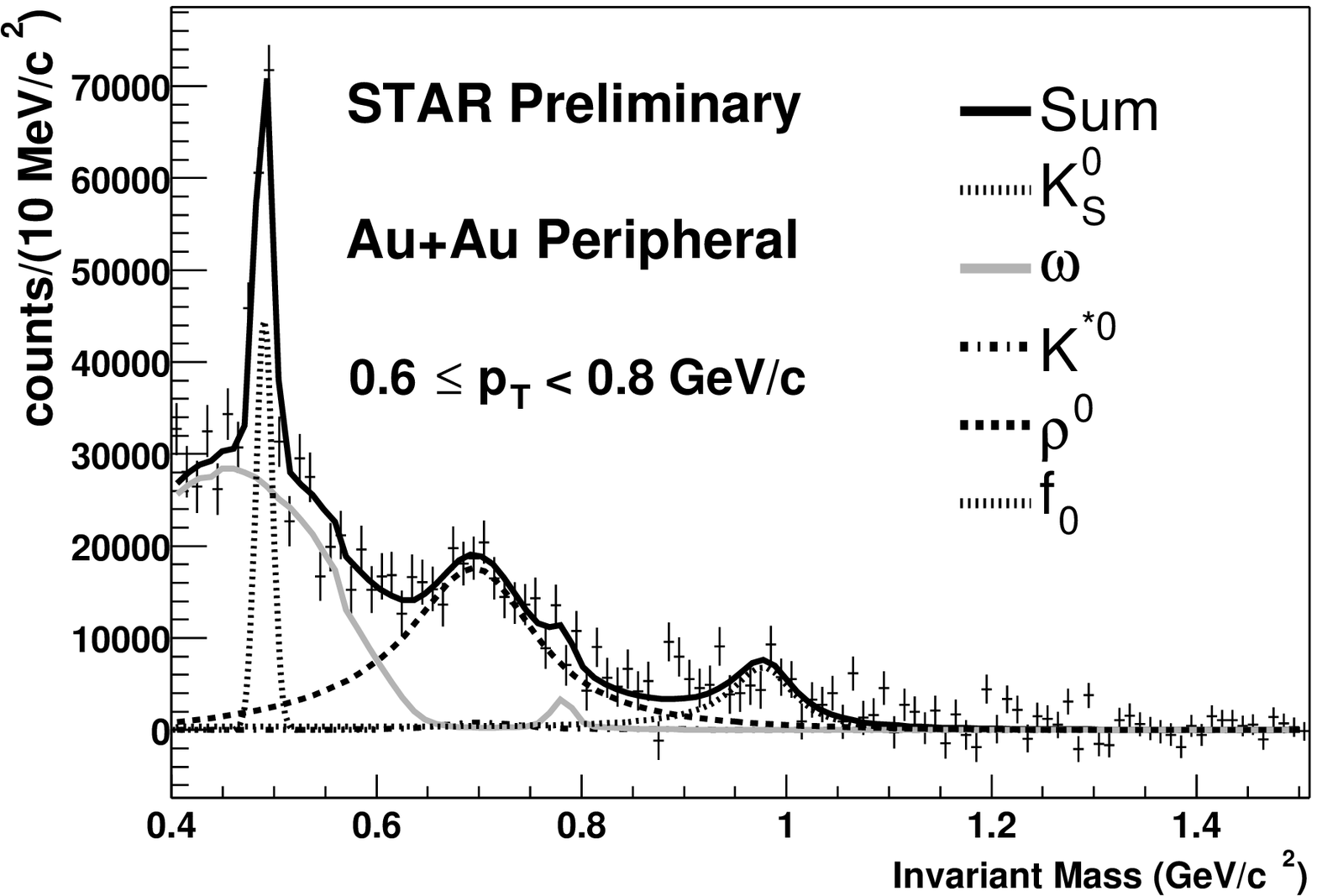}
\vspace{-0.5cm}
 \caption{The $\pi^{+} \pi^{-}$ invariant mass
distribution after like-sign background subtraction for the p+p
(left) and the Au+Au (right) interactions.}
 \label{rhoinv}
 \vspace{-0.5cm}
\end{figure}

The $\rho$ meson is a special resonance with its low mass of
770~MeV, broad width of 150 MeV and short lifetime of 1.3~fm/c.
There are several theory predictions for medium effects on the
mass and the width for the leptonic decay channels. STAR is
capable of measuring the mass and width of the $\rho$ meson (in a
corresponding "cocktail" plot) from the hadronic decay channel in
p+p and Au+Au collisions (see figure~\ref{rhoinv}). The width of
160~MeV (nominal width + detector resolution) is in agreement
with the data \cite{fat03}. A mass shift of 40~MeV
of the $\rho$ meson from p+p to Au+Au interactions has been
observed. There is also a mass shift in p+p interactions, it is
not clear if this shift scales with the Au+Au interactions. This
is still under discussion. E.V. Shuryak gave a possible
theoretical description of the relative mass shift at this
workshop so we won't discuss this topic further here
 (see \cite{shu02,shu03}). The extracted $\rho$/$\pi$ yield for p+p and
Au+Au interactions are consistent within the statistical errors.

\section{Strange Resonances}\label{data}

\begin{figure}[htb]
\centering
\includegraphics[width=0.48\textwidth]{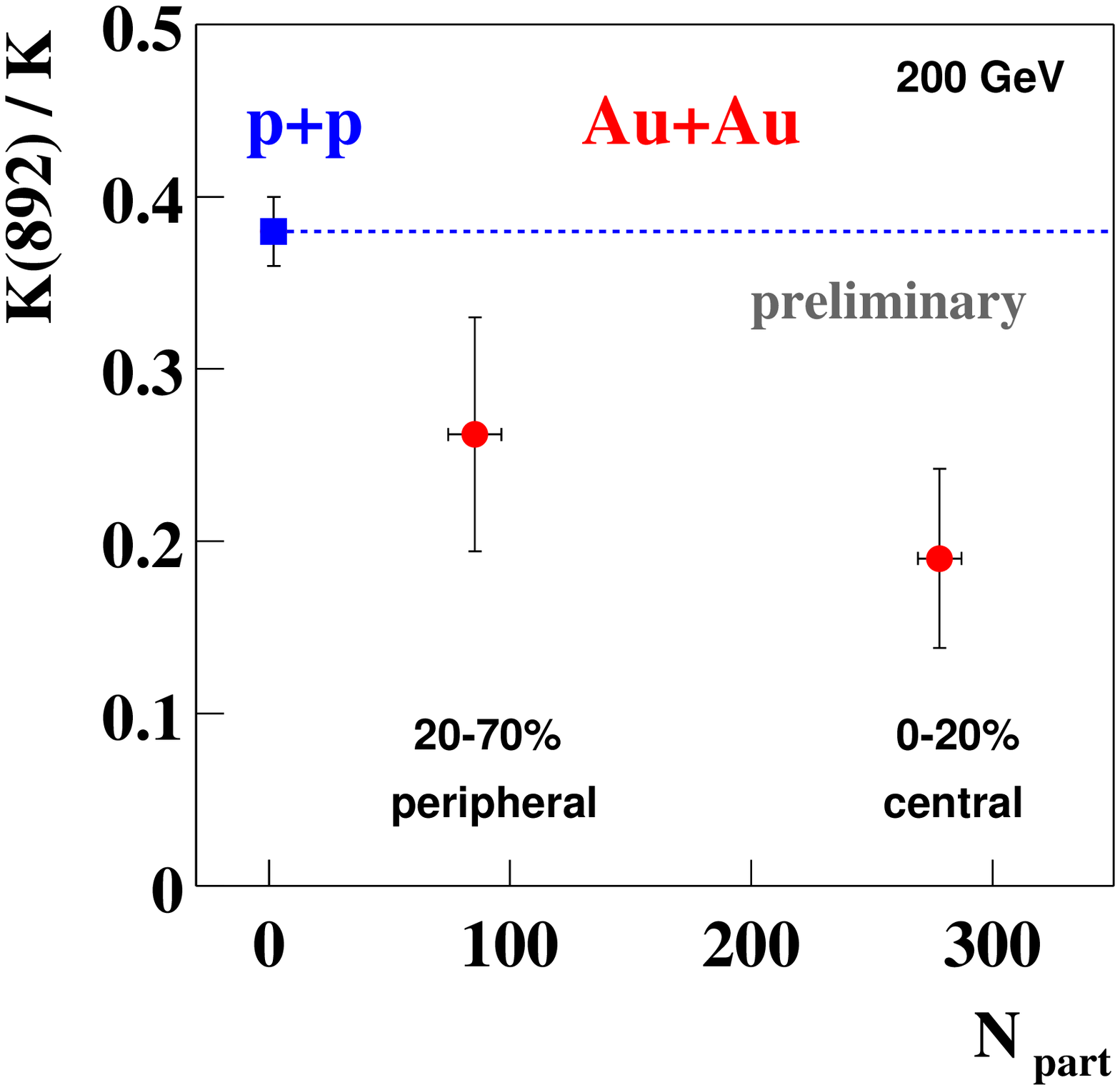}
\includegraphics[width=0.48\textwidth]{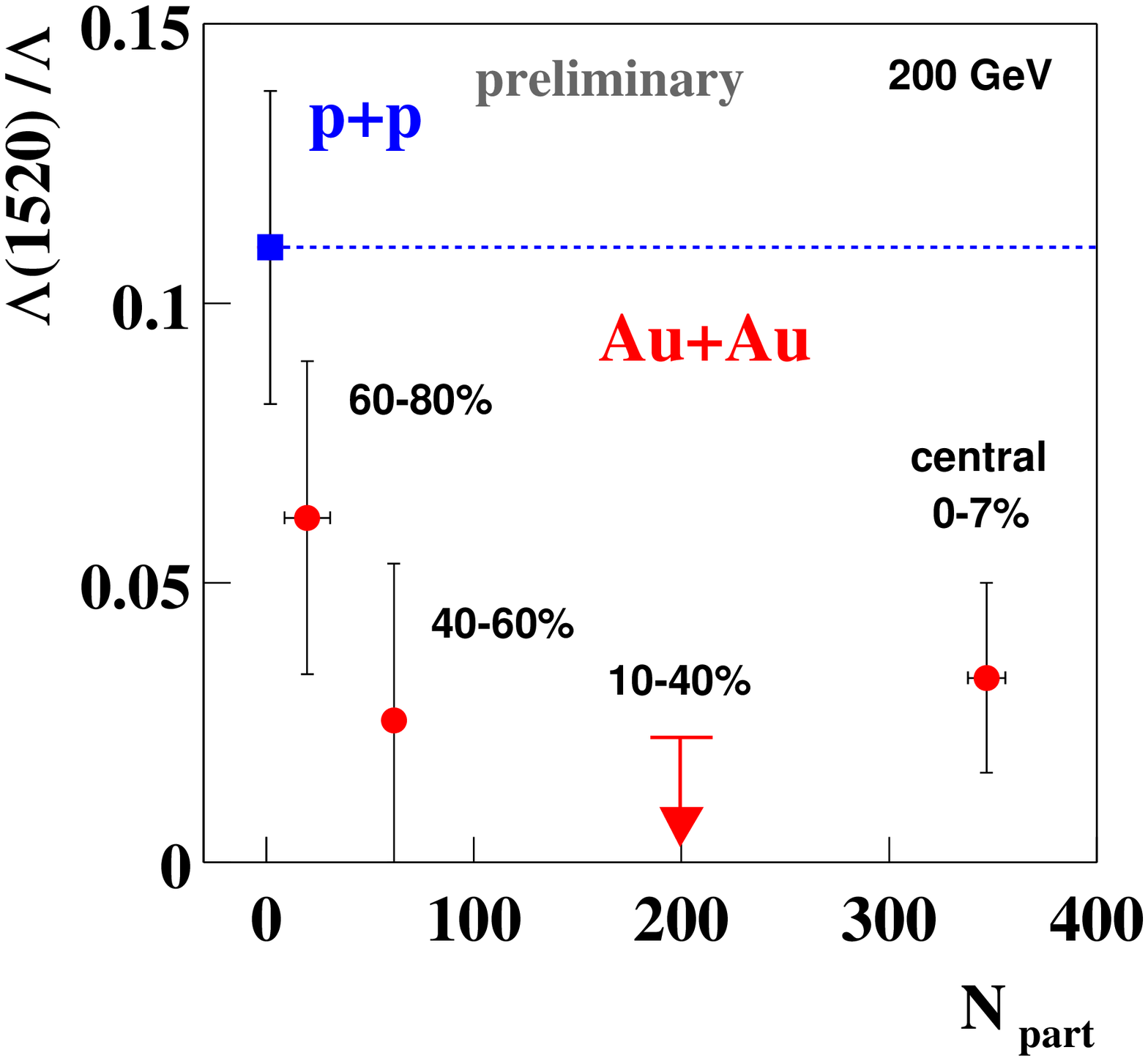}
\vspace{-0.5cm}
 \caption{Resonance over non-resonance ratios (left:
K*(892)/K \cite{zha03}, right: $\Lambda$(1520)/$\Lambda$) for p+p
and Au+Au collisions at $\sqrt{s_{\rm NN}} = $ 200 GeV, peripheral
Au+Au data from \cite{gau03}.}

 \label{part}
 \vspace{-0.5cm}
\end{figure}

The masses and the experimental widths of the strange resonances
K*(892), $\Sigma$*(1385) and $\Lambda$(1520) over the phase space
integrated invariant mass spectrum in p+p collisions are
compatible with the values from the PDG \cite{pdg98} and the
expected width coming from the momentum resolution of the
detector. Anti-particle over particle ratios of
$\bar{\Lambda}$(1520)/$\Lambda$(1520)~=~0.84~$\pm$~0.32 (stat+sys)
and $\bar{\Lambda}$/$\Lambda$~=~0.81~$\pm$~0.01 (stat) are the in
agreement ($\Lambda$ taken from \cite{bil03}) within errors. For a
comparison of resonance production in different collision systems,
the resonance over non-resonance particle ratio will take care of
the volume and the energy normalization. Figure~\ref{part} shows
the K*(892)/K and the $\Lambda$(1520)/$\Lambda$ ratio for p+p and
Au+Au collisions as a function of the number of participants. The
ratio decreases from p+p to extended Au+Au collision systems. This
behavior shows that the resonance production in Au+Au is not a
superposition of p+p interactions. This is an indication that the
surrounding extended medium of a Au+Au collision has an influence
on the resonance and/or their decay particles. Data from NA49
experiment, at an energy which is a factor $\sim$~10 lower
($\sqrt{s} = $ 17 GeV) show that the $\Lambda$(1520)/$\Lambda$ in
Pb+Pb collisions are a factor of 2 lower than in p+p collisions
\cite{fri01}. The K*(892)/K ratio at $\sqrt{s_{\rm NN}} = $
200 GeV, shown in figure~\ref{part}, decreases from p+p to more
central Au+Au collisions. This can be interpreted as direct
influence of increasing the volume of the medium. The resonance
yield relative to other particles is smaller in heavy ion
collisions than in p+p collisions. The question remains whether
the initial resonance yield is different changed and/or whether
the decay particles taken to reconstruct the resonance have been
affected by the medium.

\section{Rescattering}\label{data}

\begin{figure}[htb]
\centering
\includegraphics[width=0.60\textwidth]{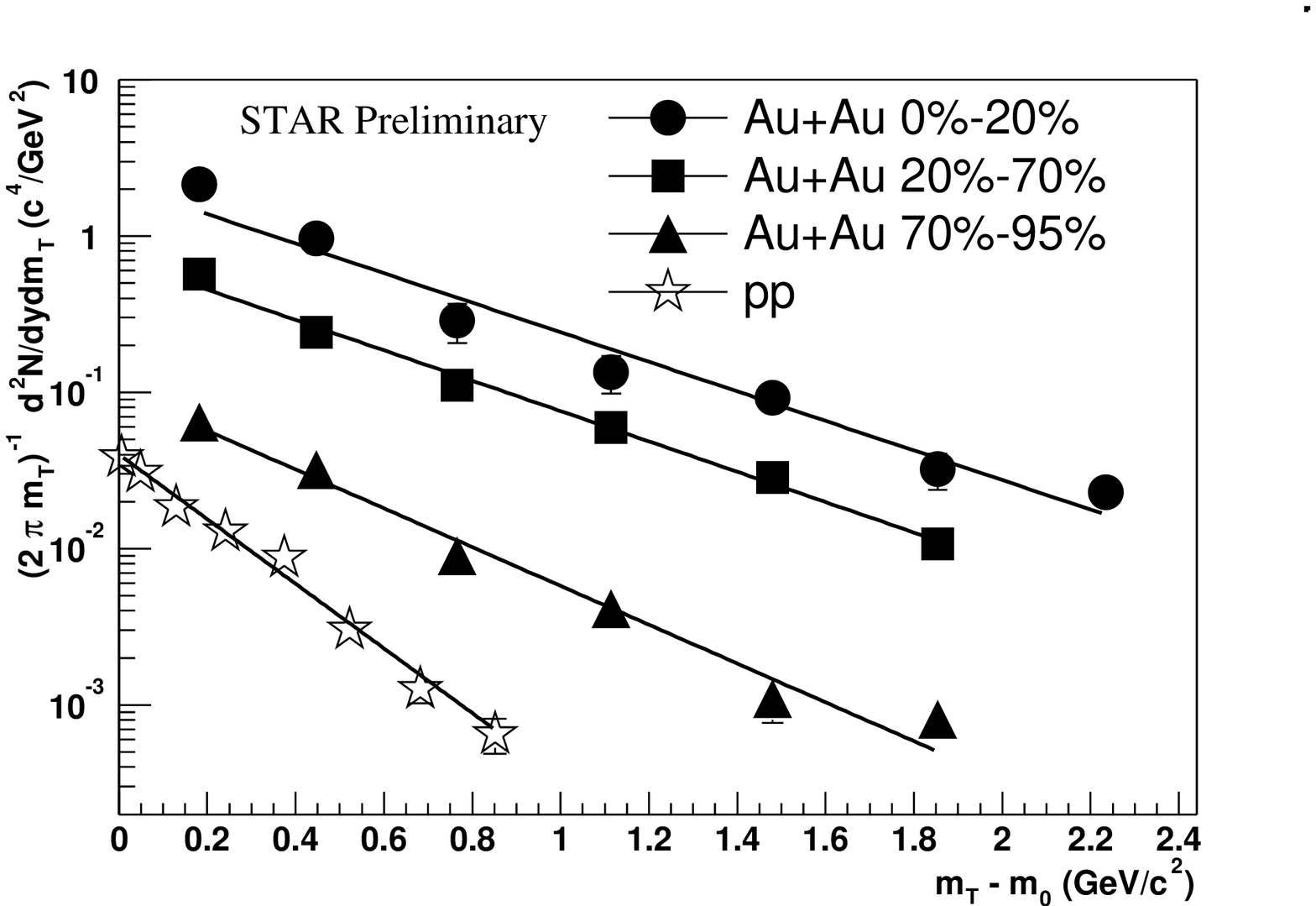}
\vspace{-0.5cm}
 \caption{Transverse mass distribution at
mid-rapidity for (K*(892)+$\overline{{\rm K}^{*}(892)})/2$ from Au+Au
an p+p interactions with statistical errors \cite{zha03}.}
\vspace{-0.5cm}
 \label{kstarslope}
\end{figure}

UrQMD predicts a signal loss in the low transverse momentum region
\cite{ble02,ble02b} due to rescattering of the decay daughters in
the medium, leading to a larger inverse slope. The inverse slope
for the K*(892) in p+p collisions is T~=~210~$\pm$~8~MeV and it
increases to T~=~350~$\pm$~23~MeV in the 70-80\% most peripheral
Au+Au collisions (see figure~\ref{kstarslope}) \cite{zha03}.
While the inverse slope for the Kaons in p+p and in the 70-80\%
most peripheral Au+Au collisions, exhibits little difference
\cite{bar02} (see table~\ref{slopes}). Thus, an increase is
observed in the slope of the K*(892) and not for the Kaons when
going from p+p interactions (with no medium) to an extended medium
in Au+Au collisions, even for the most peripheral collisions which
represent the smallest medium. This behavior was predicted by
UrQMD \cite{ble02,ble02b}. It would also be interesting if UrQMD
sees an increase of the inverse slope going from peripheral to
more central collisions.


\label{slopes}
\begin{table}[htb]
\vspace*{-12pt} \caption[]{Inverse slope parameters of transverse
mass distributions for K*(892) and K$^{-}$ for $\sqrt{s_{\rm NN}}
= $ 200 GeV p+p and Au+Au interactions \cite{fat02,zha03,bar02}}
\label{tab1} \vspace*{-14pt}
\begin{center}
\begin{tabular}{lll}
\hline\\[-10pt]
Centrality & Inverse Slope K*(892) [MeV]&  Inverse Slope K$^{-}$ [MeV]\\
\hline\\[-10pt]
0\% - 20\% Au+Au& 459 $\pm$ 32 &  321 $\pm$ 8 \\
70\% - $\sim$ 80\% Au+Au & 350 $\pm$ 23 &  202 $\pm$ 7 \\
pp & 210 $\pm$ 8 &  182 $\pm$ 5 \\
\hline
\end{tabular}
\end{center}
\end{table}

 \vspace{-0.5cm}

\section{Time Scale}\label{data}

According to thermal model predictions \cite{pbm01} (becattini)
the production yield of the K*(892)/K in heavy ion collisions at
$\sqrt{s_{\rm NN}} = $ 200 GeV is within the errors a factor of 2
smaller than the predicted value. Due to the large errors for the
K*(892) yield in the $\sqrt{s_{\rm NN}} = $ 130 GeV data we were
not previously sensitive to this ratio discrepancy with the
thermal model predictions \cite{adl02}. Also the first
$\Lambda$(1520) yield measurement show a 50\% lower value than
that predicted by thermal models. Model calculations that include
particle yields from thermal production and a lifetime for the
system after chemical freeze-out, where the cross section of
rescattering but not regeneration for the decay daughters is
included, predict a signal loss for the measured resonances. Using
the measured values of K(892)/K and $\Lambda$(1520)/$\Lambda$, and
a chemical freeze out temperature of 175 MeV the lifetime between
chemical and kinetic freeze out is estimated to be 4-6 fm/c
\cite{tor01,tor01a,mar02}. Additional measurements from particle
correlations gives 9-10 fm/c for the total lifetime in rough
agreement with this number \cite{lis03,pan03,nig03}.

\section*{Acknowledgements}
I would like to thank the organizers who invited me to such a well
organized workshop which left enough time for detailed and long
discussions. I also would like to thank the STAR collaboration for
support in presenting this data. My research is supported by the
Humboldt Foundation, Germany.

\vfill\eject
\end{document}